\documentclass[12pt,a4paper]{article}
\usepackage[colorlinks]{hyperref}
\usepackage{amsmath,amssymb}
\usepackage{url,color}
\usepackage{graphicx}
\usepackage{array,dsfont}
\usepackage{float}
\usepackage{booktabs,tabularx}
\usepackage[justification=justified,singlelinecheck=false]{caption,subfig}
\usepackage{placeins}
\newcolumntype{M}[1]{>{\centering\arraybackslash}m{#1}}
\newcolumntype{N}{@{}m{0pt}@{}}
\makeatletter
\newif\if@preliminary
\@preliminaryfalse
\def\preliminary{\@preliminarytrue}

\newcommand{\vH}{\mathbf{H}}

\newcommand{\vD}{\mathbf{D}}

\newcommand{\bss}{\begin{tiny}}
\newcommand{\ess}{\end{tiny}}

\def\preprintno#1{\def\@preprintno{#1}}
\def\address#1{\def\@address{#1}}
\def\email#1#2{\thanks{\tt #1@{}#2}}
\def\abstract#1{\def\@abstract{#1}}
\renewcommand\abstractname{ABSTRACT}
\newlength\preprintnoskip
\newlength\abstractwidth
\@titlepagetrue
\renewcommand\maketitle{\begin{titlepage}%
  \let\footnotesize\small
  \hfill\parbox{\preprintnoskip}{%
  \begin{flushright}\@preprintno\end{flushright}}\hspace*{1cm}
  \vskip 60\p@
  \begin{center}%
    {\Large\bf\boldmath \@title \par}\vskip 1cm%
    {\sc\@author \par}\vskip 3mm%
    {\@address \par}%
    \if@preliminary
      \vskip 2cm {\large\sf PRELIMINARY DRAFT \par \@date}%
    \fi
  \end{center}\par
  \@thanks
  \vfill
  \begin{center}%
    \parbox{\abstractwidth}{\centerline{\abstractname}%
    \vskip 3mm%
    \@abstract}
  \end{center}
  \end{titlepage}%
  \setcounter{footnote}{0}%
  \let\thanks\relax\let\maketitle\relax
  \gdef\@thanks{}\gdef\@author{}\gdef\@address{}%
  \gdef\@title{}\gdef\@abstract{}\gdef\@preprintno{}
}%
\def\@citex[#1]#2{\if@filesw\immediate\write\@auxout{\string\citation{#2}}\fi
  \def\@citea{}\@cite{\@for\@citeb:=#2\do
    {\@citea\def\@citea{,\penalty\@m}\@ifundefined
       {b@\@citeb}{{\bf ?}\@warning
       {Citation `\@citeb' on page \thepage \space undefined}}%
\hbox{\csname b@\@citeb\endcsname}}}{#1}}
\def\citerange{\@ifnextchar [{\@tempswatrue\@citexr}{\@tempswafalse\@citexr[]}}
\def\@citexr[#1]#2{\if@filesw\immediate\write\@auxout{\string\citation{#2}}\fi
  \def\@citea{}\@cite{\@for\@citeb:=#2\do
    {\@citea\def\@citea{--\penalty\@m}\@ifundefined
       {b@\@citeb}{{\bf ?}\@warning
       {Citation `\@citeb' on page \thepage \space undefined}}%
\hbox{\csname b@\@citeb\endcsname}}}{#1}}
\long\def\@makecaption#1#2{%
  \sbox\@tempboxa{#1: \emph{#2}}%
  \ifdim \wd\@tempboxa >\hsize
    #1: \emph{#2}\par
  \else
    \hbox to\hsize{\hfil\box\@tempboxa\hfil}%
  \fi
  \vskip\belowcaptionskip}
\def\fmslash{\@ifnextchar[{\fmsl@sh}{\fmsl@sh[0mu]}}
\def\fmsl@sh[#1]#2{%
  \mathchoice
    {\@fmsl@sh\displaystyle{#1}{#2}}%
    {\@fmsl@sh\textstyle{#1}{#2}}%
    {\@fmsl@sh\scriptstyle{#1}{#2}}%
    {\@fmsl@sh\scriptscriptstyle{#1}{#2}}}
\def\@fmsl@sh#1#2#3{\m@th\ooalign{$\hfil#1\mkern#2/\hfil$\crcr$#1#3$}}
\makeatother

\newcommand\ltap{\
  \raise.3ex\hbox{$<$\kern-.75em\lower1ex\hbox{$\sim$}}\ }
\newcommand\gtap{\
  \raise.3ex\hbox{$>$\kern-.75em\lower1ex\hbox{$\sim$}}\ }

\newcommand\simge{\mathrel{%
   \rlap{\raise 0.511ex \hbox{$>$}}{\lower 0.511ex \hbox{$\sim$}}}}
\newcommand\simle{\mathrel{
   \rlap{\raise 0.511ex \hbox{$<$}}{\lower 0.511ex \hbox{$\sim$}}}}

\newcommand\be{\begin{equation}}
\newcommand\ee{\end{equation}}
\newcommand\bea{\begin{eqnarray}}
\newcommand\eea{\end{eqnarray}}
\newcommand\ba{\begin{array}}
\newcommand\ea{\end{array}}

\def\bq{\begin{equation}}
\def\eq{\end{equation}}
\def\ba{\begin{eqnarray}}
\def\ea{\end{eqnarray}}

\newcommand{\LL}{\mathcal{L}}
\newcommand{\tr}{\mathop{\rm tr}}

\newcommand{\ii}{{\rm i}}

\begin{document}

\date{\today}

\preprintno{1606.xxxxx [hep-ph]; DESY 16-097, SI-HEP-2016-16}

\title{New Physics/Resonances in Vector Boson Scattering at the LHC}

\author{\underline{J\"urgen Reuter}\email{juergen.reuter}{desy.de}$^a$, 
  Wolfgang Kilian\email{kilian}{physik.uni-siegen.de}$^b$,
  Thorsten Ohl\email{ohl}{physik.uni-wuerzburg.de}$^c$,
  Marco Sekulla\email{marco.sekulla}{kit.edu}$^d$}

\address{\it%
$^a$DESY Theory Group, 
D-22603 Hamburg, Germany
\\[.5\baselineskip]
$^b$
Department of Physics, %
University of Siegen, %
D--57068 Siegen, Germany
\\[.5\baselineskip]
$^c$
Faculty of Physics and Astronomy, %
W\"urzburg University, %
D--97074 W\"urzburg, Germany
\\[.5\baselineskip]
$^d$
Institute for Theoretical Physics, %
Karlsruhe Institute of Technology, %
D--76128 Karlsruhe, Germany
\\[4\baselineskip]
Talk given at the "Conference on New Physics at the Large Hadron
Collider", 29.02. - 04.03.2016, Nanyang University, Singapore 
}

\abstract{
Vector boson scattering is (together with the production of multiple
electroweak gauge bosons) the key process in the current run 2 of LHC
to probe the microscopic nature of electroweak symmetry
breaking. Deviations from the Standard Model are generically
parameterized by higher-dimensional operators, however, there is a
subtle issue of perturbative unitarity for such approaches for the
process above. We discuss a parameter-free unitarization prescription
to get physically meaningful predictions. In the second part, we
construct simplified models for generic new resonances that can appear
in vector boson scattering, with a special focus on the technicalities
of tensor resonances.
}

\maketitle


\section{Motivation}
Run I of the LHC has not only revealed a Standard Model-like Higgs
boson~\cite{Aad:2012tfa,Chatrchyan:2012xdj} together with measuring
its mass and some of its properties and couplings, 
but also established the scattering process of electroweak gauge
bosons\cite{Aad:2014zda,ATLAS:2014rwa,CMS:2015jaa} (VBS) as
predicted by the Standard Model (SM). This process gives insights in
the nature of the electroweak symmetry breaking (EWSB) sector and
further fundamental properties of the Higgs. In the SM, the
electroweak breaking sector is described as a weakly interacting
theory, where the Higgs boson is strongly suppressing the vector
boson scattering process at high center-of-mass energies and the
scattering amplitude is dominated by the transversal vector
boson scattering.

Without the Higgs the VBS scattering amplitudes $VV \rightarrow VV$,
where $V$ is $W^\pm,Z$, will rise with $s / v^2$ due to the
dominant contribution of scalar Goldstone-boson scattering,
which represents the longitudinal degrees of freedom of the vector
boson scattering. The electroweak interactions would become strongly
interacting in the TeV range. However, the initial limits on VBS are
rather weak and only scales close to the pair-production threshold
of $\sim 200$ GeV are probed. Run II and III of the LHC and future
(high-energy) $e^+e^-$ colliders will improve the accuracy 
and provide new insights in the origin of EWSB. The delicate
cancellation between the EW gauge bosons and the Higgs boson in VBS,
makes this channel an ideal, though not easy place to search for new
physics. 

The discussion in these proceedings is based on our publications
in~\cite{Alboteanu:2008my,Kilian:2014zja,Kilian:2015opv,Sekulla:2015}.


\section{Effective Field Theory, Perturbative Unitarity and Unitarization}

To study new physics in the VBS process generically, we
will use the framework of Effective Field Theories (EFT). A set of
higher dimensional operators extends the SM Lagrangian to quantify
deviations from the SM, which originate from some new physics at a
high energy scale $\Lambda_i$ as 
\begin{align}
  \LL= \LL_{SM} + \sum_i \frac{C_i}{\Lambda_i^{d-4}} \mathcal{O}^{d}.
\end{align}
Here, $C_i$ are the associated Wilson coefficients of the operators.
Due to lack of knowledge of both parameters, we introduce the ratio
coupling $F_i=\frac{C_i}{\Lambda_i^{d-4}}$. 

Many different operator bases have been proposed for the electroweak
sector, an overview and also translations between them have been
discussed e.g. in~\cite{Baak:2013fwa,Degrande:2013rea}. Here, for
illustrative purposes, we take just two different operators,
$\mathcal{O}_{HD}$ as a dim-6 operator, and the two dim-8 operators,
$\mathcal{O}_{S,0}$ and $\mathcal{O}_{S,1}$. All of these operators
could arise easily in popular scenarios of new physics beyond the SM
(BSM) like Composite Higgs, Little Higgs or Extra Dimensions. The LHC
experiments are studying all of three of them to gain sensitivity in
various channels like dibosons, tribosons, precision Higgs data and
VBS. The operators are given by
\begin{alignat}{3}
  \label{LL-HD}
  \LL_{HD} &=
  & & F_{HD}\ &&
  \tr{{\vH^\dagger\vH}- \frac{v^2}{4}}\cdot
  \tr{\left (\vD_\mu \vH \right )^\dagger \left (\vD^\mu \vH \right )} \, ,
  \\
  \label{LL-S0}
  \LL_{S,0}&=
  & &F_{S,0}\ &&
  \tr{ \left ( \vD_\mu \vH \right )^\dagger \vD_\nu \vH}
  \cdot \tr{ \left ( \vD^\mu \vH \right )^\dagger \vD^\nu \vH} \, ,
  \\
  \label{LL-S1}
  \LL_{S,1}&=
  & &F_{S,1}\ &&
  \tr{ \left ( \vD_\mu \vH \right )^\dagger \vD^\mu \vH}
  \cdot \tr{ \left ( \vD_\nu \vH \right )^\dagger \vD^\nu \vH} \, .
\end{alignat}

Due to the unknown microscopic picture of the underlying energy giving
rise to these operators, the validity range of the EFT is also apriori
unknown. In this case, the unitarity condition is used to determine the
validity of the EFT.

\begin{figure}[htb]
  \includegraphics[width=0.5\linewidth]{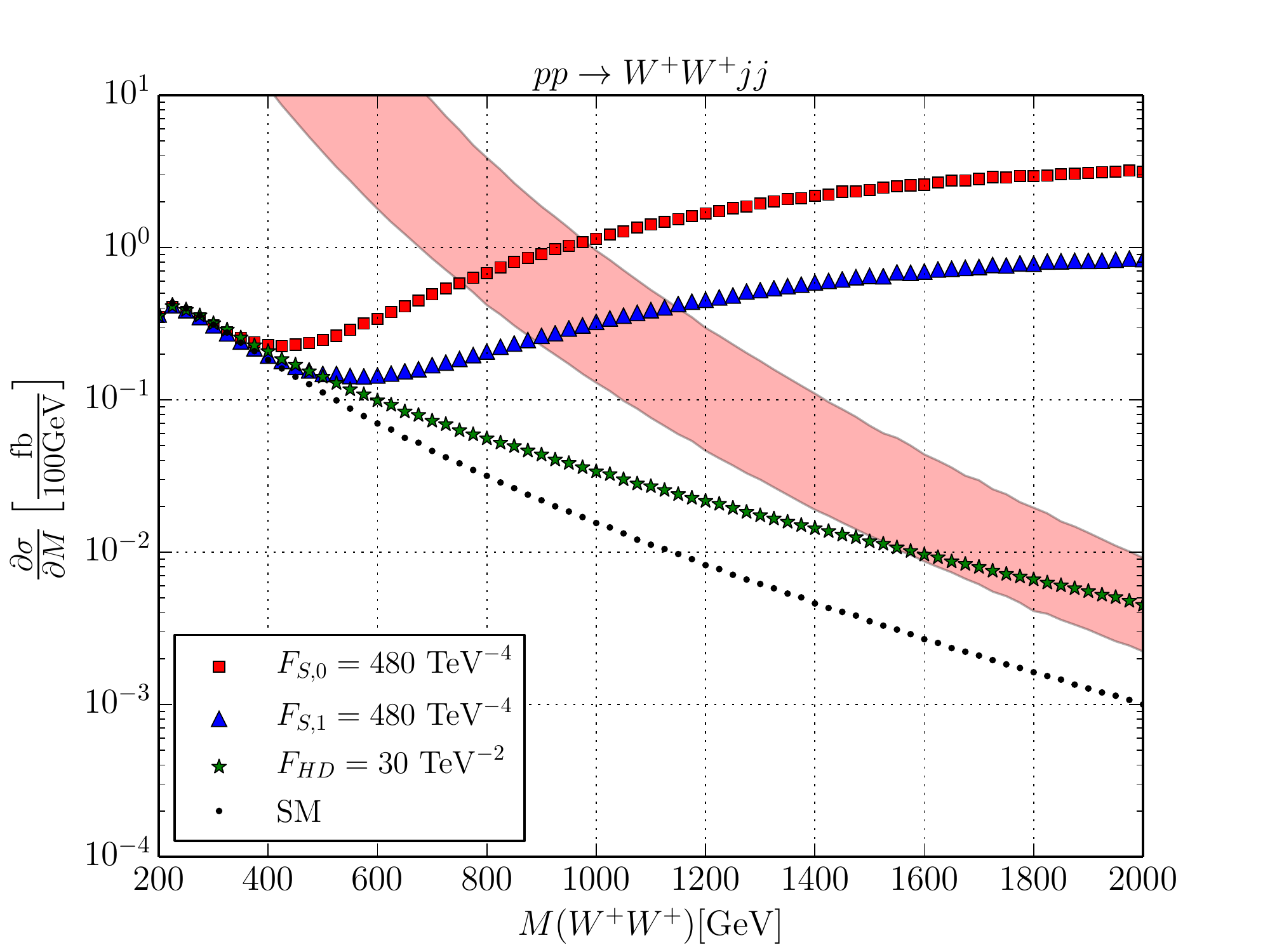}
  \includegraphics[width=0.5\linewidth]{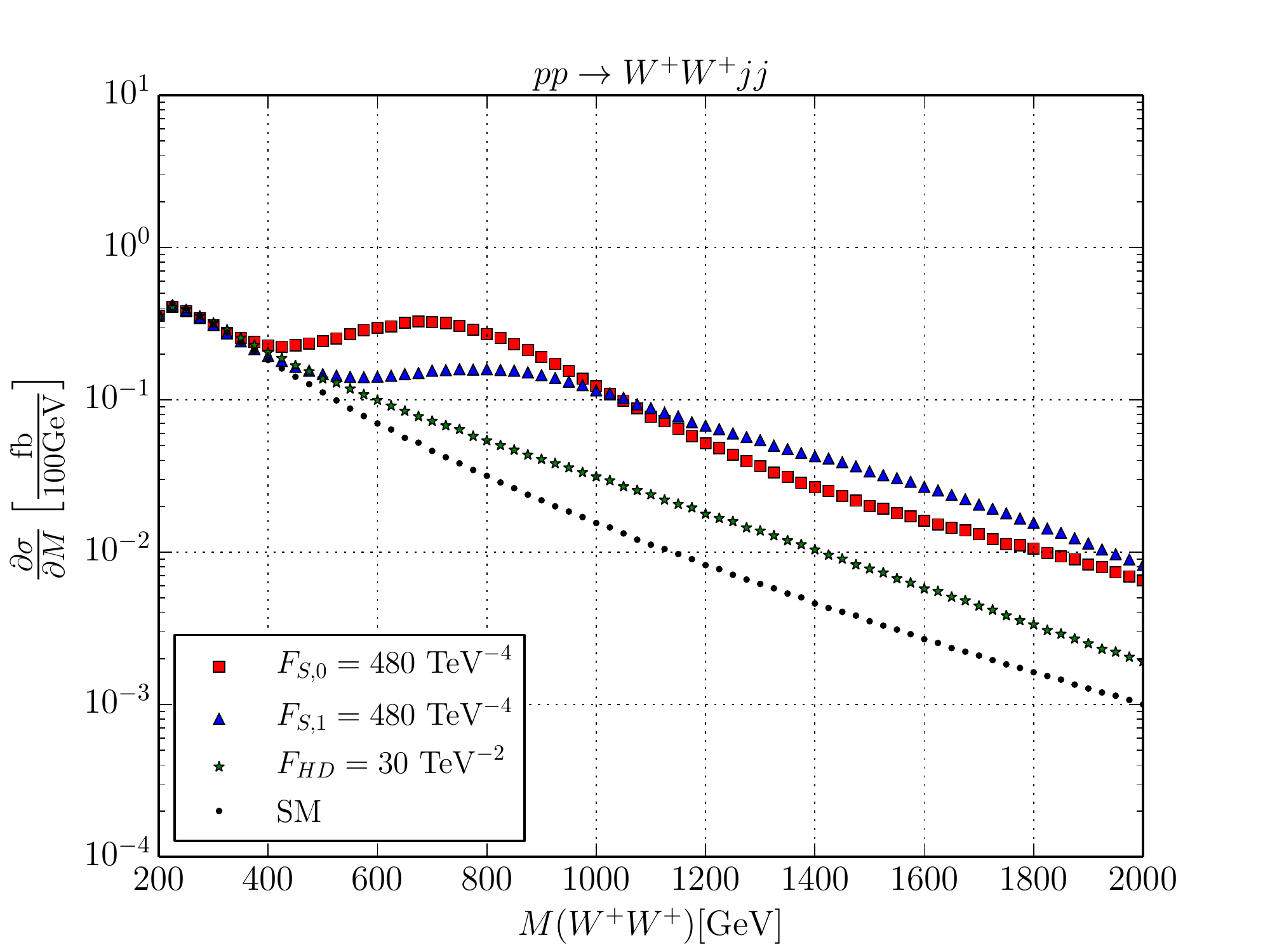}
  \caption{\label{fig:ww_unit}
    $pp\to W^+W^+jj$, left: naive EFT results that violate
    unitarity, QCD contributions neglected. The band describes maximal
    allowed values, due to unitarity constraints, for the
    differential cross section. The lower bound describes the
    saturation of $\mathcal{A}_{20}$ and the upper bound describes the
    simultaneous saturation of $\mathcal{A}_{20}$ and
    $\mathcal{A}_{22}$, right : unitarized result.
    Cuts: $M_{jj} > 500$ GeV;
    $\Delta\eta_{jj} > 2.4$;
    $p^j_T > 20$ GeV;
    $|\eta_j| > 4.5$.}
\end{figure}

In the left-hand side of Fig.~\ref{fig:ww_unit}, the cross section for
the complete LHC process $pp\rightarrow W^+W^+jj$ at leading order --
computed using the Monte-Carlo generator \texttt{WHIZARD} with
CTEQ6L PDF sets -- is shown. The SM curve is compared to three curves
for models which contain a single nonzero coefficient for the
three different effective higher dimensional operators, respectively.
For an indication of the unitarity limits, we have included a
quartic Goldstone interaction amplitude with a constant coefficient
$a_{IJ}=\ii$ in the $I=2$ and $J=0,2$ isospin and spin channels and
recomputed the process with this modification. The Goldstone boson
scattering amplitudes are very good approximations to the scattering
of longitudinal EW gauge bosons by means of the Goldstone boson
equivalence theorem. By projecting the partial waves into their spin
and isospin components, the optical theorem is used to determine the
condition for perturbative unitarity in the same way as
in~\cite{Lee:1977yc}. Further details are listed in
\cite{Kilian:2014zja}. At high invariant mass $M_{VV}$  of the
$WW$-scattering system, the enhancement the crossection by
$\frac{M_{WW}^8}{m_H^8}$ in comparison to the SM due to the dimension
eight operators are dominant. The coefficients of the higher dimension
operators are chosen within current LHC bounds. We concentrated to the
like-sign $WW$ scattering as this is the clearest channel at the LHC
with the smallest backgrounds. It only appears in the isospin two
channel. In the light red band, we plotted the unitarity limit by
demanding that the $\mathcal{A}_{20}$ and $\mathcal{A}_{22}$ for
isospin two and spin zero and two, respectively, are saturated,
i.e. reaching there maximal value of $32\pi$ allowed by perturbative
unitarity. 

The prediction of the dimension eight operators violate the unitarity limit and
become unphysical in an energy regime, which can be tested at the LHC.
Naively, one could introduce a cut-off to forbid these unphysical
events manually (a prescription also partially used by ATLAS and CMS,
known as 'event clipping'). Such a cutoff could also be motivated
theoretically by the argument that these events could have never
arisen in a UV-complete theory. However, this leads to a sharp edge in
the distribution (at level of the vector bosons) which does not
resemble any sensible approximation to a UV-complete theory, and
furthermore there are also experimental constrictions for doing so:
In case of the $W^+W^+$ scattering, the final state includes two
neutrinos and the $WW$ invariant mass cannot be experimentally
reconstructed. Other methods to treat this high-energy regime are by
means of so-called form factors which, however, depend on at least two
arbitrary parameters, the exponent of the momentum dependence in the
denominator (the 'multipole' parameter) and the cutoff scale which a
\begin{figure}[hbt]
  \begin{center}
    \includegraphics[width=0.8 \textwidth]{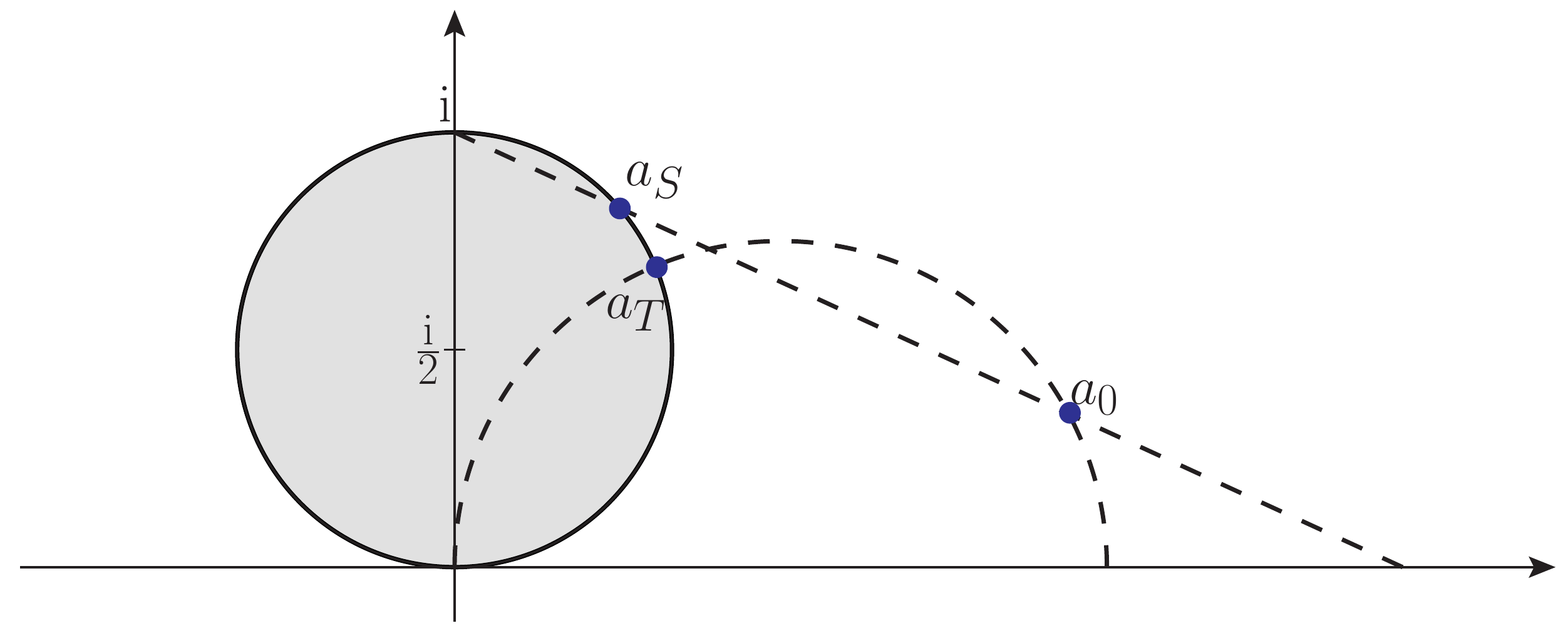}
  \end{center}
  \caption{Geometrical representation: stereographic projection vs
    Thales projection.}
  \label{fig:ThalesStereo2}
\end{figure}
priori has nothing to do with the scale $\Lambda$ appearing in front
of the Wilson coefficients. 

\begin{figure}[tb]
  \begin{center}
    \subfloat[SM]{\includegraphics[scale=0.32]{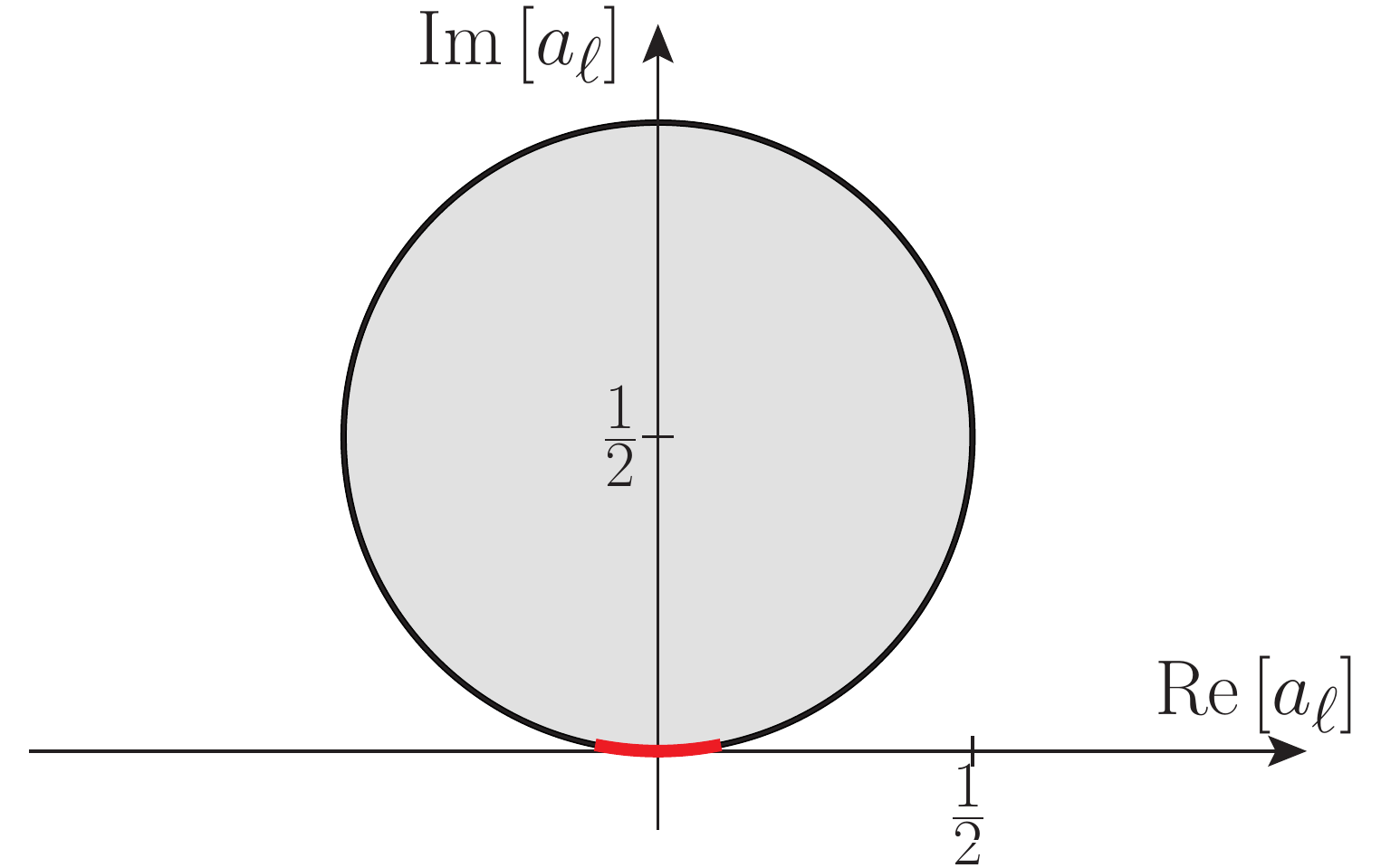}
      \label{subfig:sm}}
    \subfloat[Bare EFT, high-energy]{\includegraphics[scale=0.32]{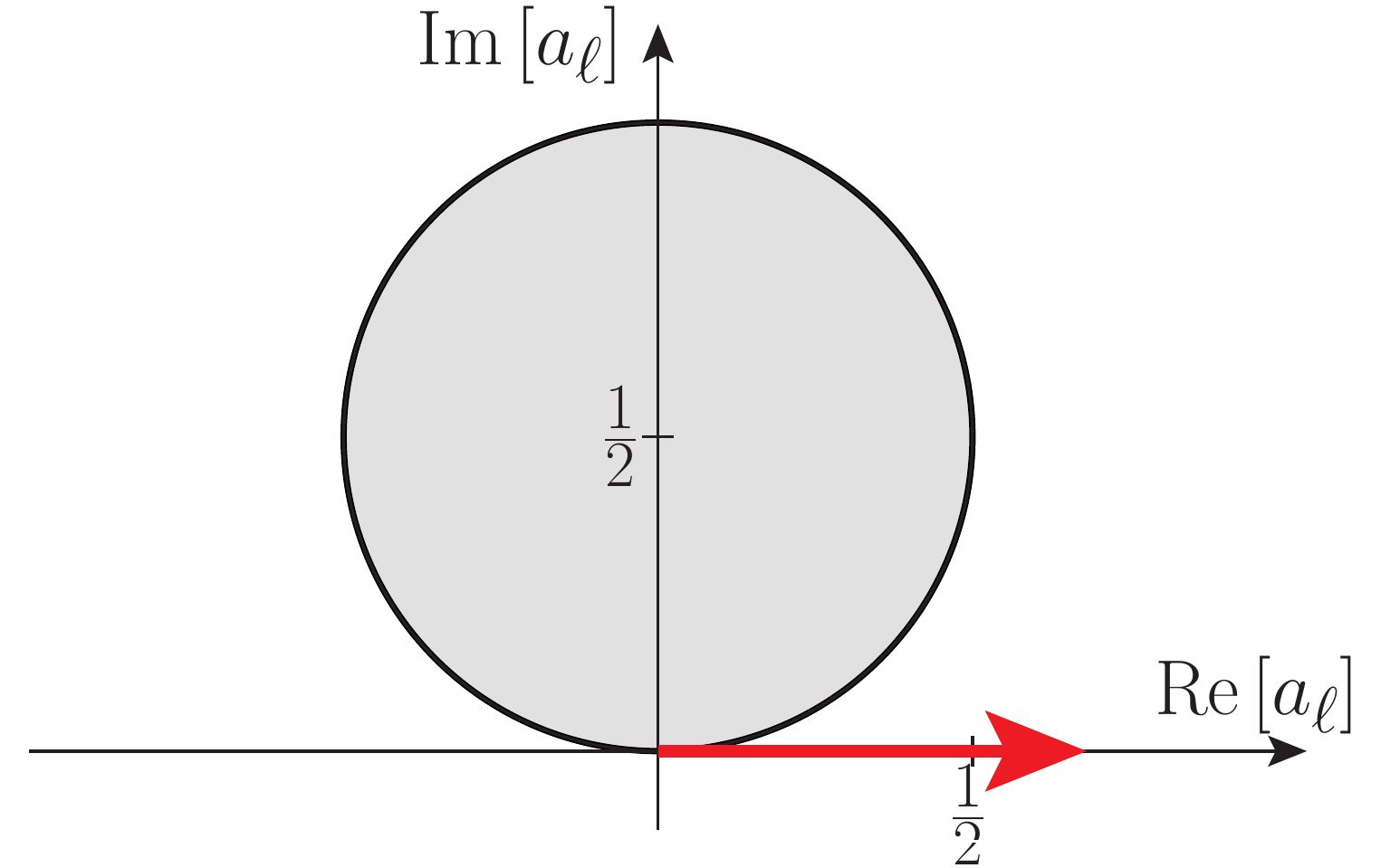}
      \label{subfig:eft}}\\
    \subfloat[Saturation]{\includegraphics[scale=0.32]{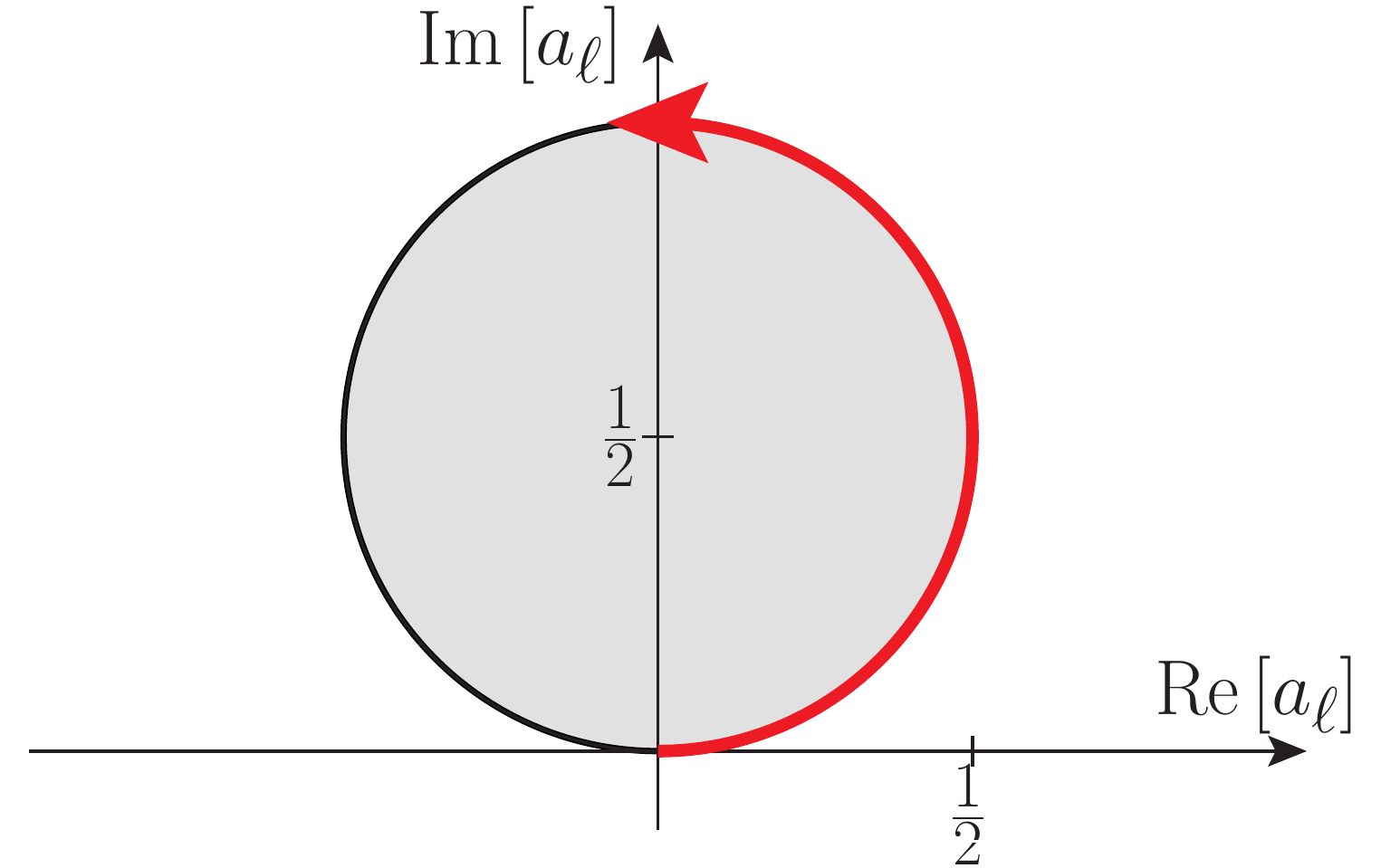}
      \label{subfig:saturation}}
      \subfloat[Inelastic channels]{\includegraphics[scale=0.32]{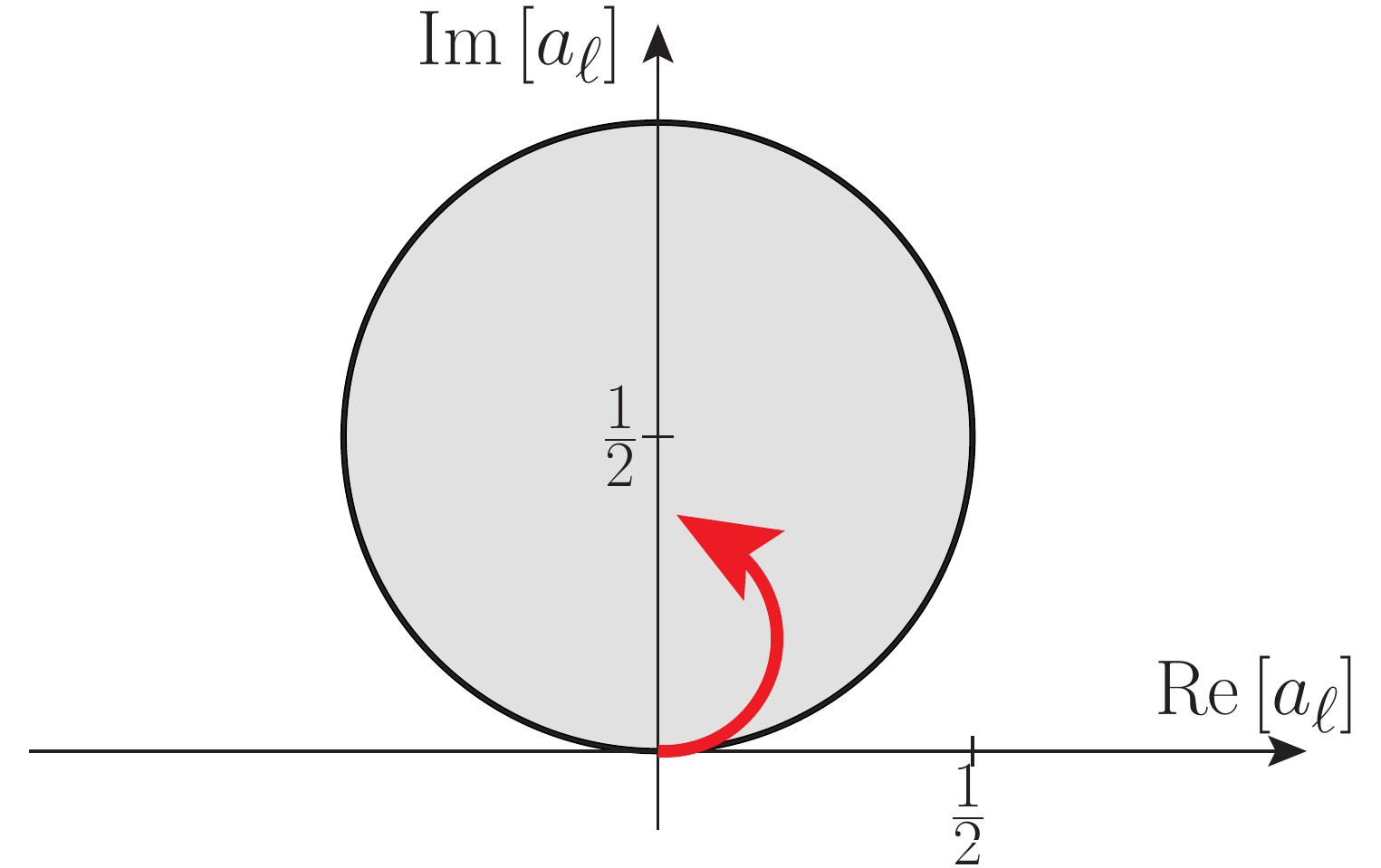}
        \label{subfig:inelastic}}
      \subfloat[Resonance]{\includegraphics[scale=0.32]{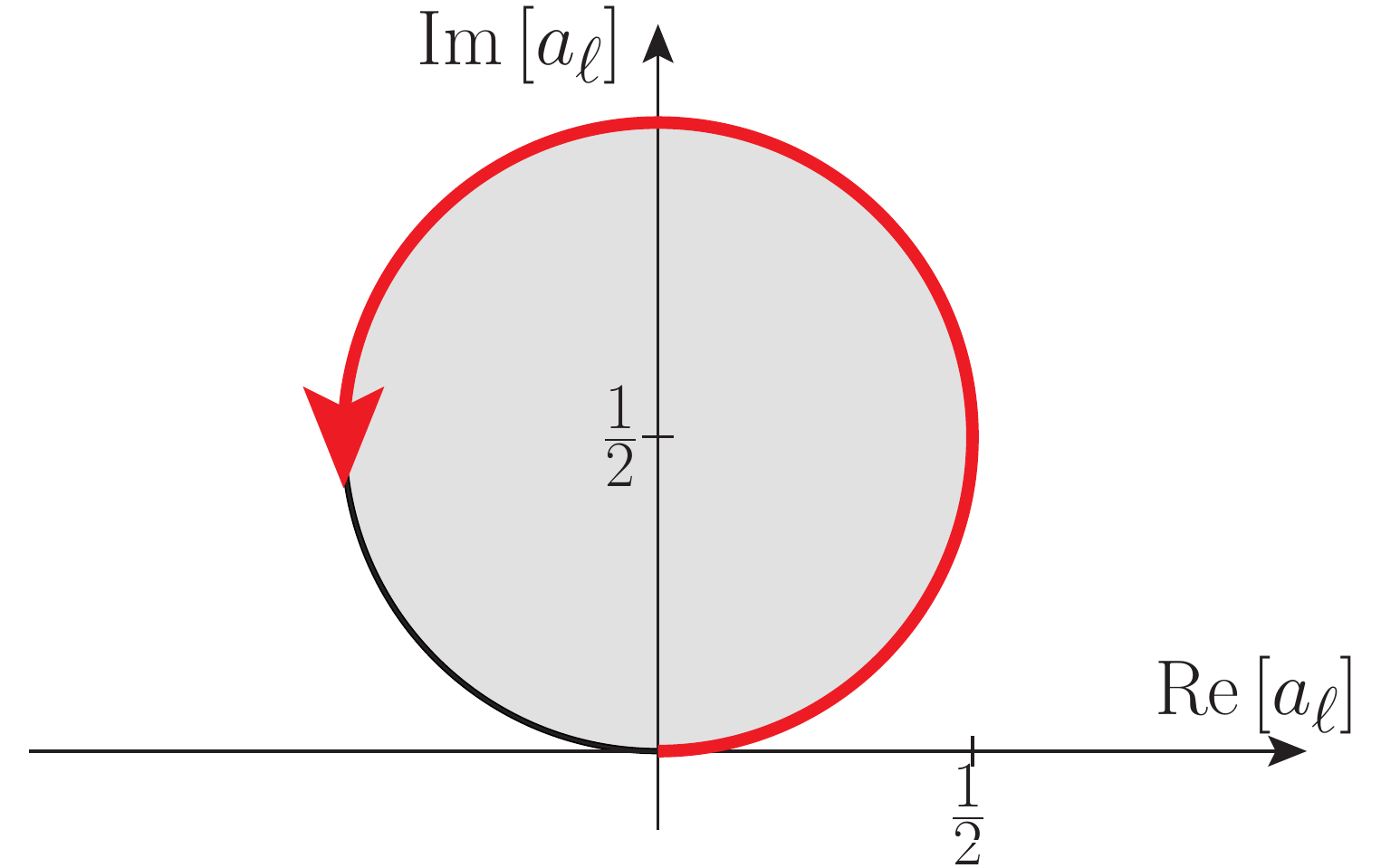}
        \label{subfig:resonance}}
  \end{center}
  \caption{Possible situations for scattering amplitudes respecting the Argand circle.}
  \label{fig:argand}
\end{figure}

In order to have a meaningful description that does not depend on any
parameters lacking physical motivation, we introduce the $T$-matrix
unitarization scheme (cf.~Fig.~\ref{fig:ThalesStereo2}) as a general
extension of the $K$-matrix unitarisation to provide event samples,
which satisfy the unitarity bound. The T-matrix scheme
is applicable for cases where the amplitude has an imaginary
part itself already, and is also defined without relying on a
perturbative expansion. For more details cf.~\cite{Kilian:2014zja}. The
right-hand side of Fig.~\ref{fig:ww_unit} shows the damping of the
cross sections for high energies due to the saturation of the
amplitudes. The T-matrix scheme is only one extrapolation for possible
high energy scenarios. All physical scenarios have to fullfil the
unitarity condition which is graphically represented by the Argand
circle. If no new physics is involved in the electroweak sector, the
elastic scattering amplitude of the Standard model will
stay at the origin on the bottom of the Argand circle
(Fig.~\ref{subfig:sm}). If the EFT is naively added, amplitudes start
to rise and will leave the Argand circle to finally
violate unitarity (cf.~Fig.~\ref{subfig:eft}, as there are new degrees
of freedom in the strict EFT, the amplitude can never develop an
imaginary part to return to the Argand circle). To remedy this
unphysical behavior of the amplitude, unitarization prescriptions are
introduced to project the amplitude back onto the Argand circle. T-matrix
unitarization saturizes the amplitude, in the sense that it is equivalent
to an infinitely broad resonance at infinity, similarly to a strongly
interacting continuum present over an extended range in momentum
space. Another option to correct the unphysical EFT prediction is
using the form-factor scheme, a possible case of entering the
inelastic regime with additional channels opening up
(cf.~Fig.~\ref{subfig:inelastic}).  A third approach would be the
addition of additional resonances (either weakly or strongly coupled), which 
could be (part of the) origin for the dim-8 operators
(cf.~Fig.~\ref{subfig:resonance}).  Here, the amplitude will ideally fall
again beyond the resonance, but could show a rise again due to
continuum contributions or the onset of a further resonance.


\section{Resonances and Simplified Models}
  
As the LHC is intended to be a discovery machine, it might be
advantageous to assume that a new resonance or particle might be
within the kinematic reach of the machine, especially given the high
amount of luminosity to be collected in runs II and III. In order to
be as general as possible in studying what kind of resonances could
show in vector boson scattering -- specific models would be Two-Higgs
double models, including the (N)MSSM, Composite Higgs, Little
Higgs~(for limits cf.~e.g.~\cite{LittleHiggs}), Twin Higgs, etc. --
we classify all resonances that can couple to the electroweak diboson
systems according to their spin and isospin quantum numbers. For
simplicity, we neglect couplings to photons, but of course they are
present due to EW gauge invariance. These possible resonances can be
categorized in terms of the approximate $SU(2)_L \times SU(2)_R$,
which is a good approximation for weak boson scattering, and the
spin. The $(0,0)$ and the $(1,1)$ representation of the $SU(2)_L
\times SU(2)_R$ are abbreviated as isoscalar and isotensor,
respectively. We can distinguish the resonances for elastic vector
boson scattering into a isoscalar scalar $\sigma$, a isoscalar tensor
$f$, a isotensor scalar $\phi$ and a isotensor tensor $X$. The
interaction with longitudinal vector bosons is modeled by the 
following currents:
\begin{subequations}
  \begin{alignat}{2}
    J_{\sigma} &= F_\sigma&&
    \tr{ \left ( \vD_\mu \vH \right )^\dagger \vD^\mu \vH} \, ,\\
    J_{\phi} &= F_\phi&&
    \left (
    \left ( \vD_\mu \vH \right )^\dagger \otimes \vD^\mu \vH
    +\frac {1}{8} \tr{\left ( \vD_\mu \vH \right )^\dagger \vD^\mu \vH }
    \right )\tau^{aa} \, , \\
    J^{\mu \nu}_f&=
    F_f &&\left (
    \tr{ \left ( \vD^\mu \vH \right )^\dagger \vD^\nu \vH}
    - \frac{c_f}{4} g^{\mu \nu}
    \tr{ \left ( \vD_\rho \vH \right )^\dagger \vD^\rho \vH}
    \right)  \, , \\
    J^{\mu \nu}_{X }&=
    F_X& & \Bigg [
      \frac{1}{2} \left (
      \left ( \vD^\mu \vH \right )^\dagger \otimes \vD^\nu \vH
      + \left (  \vD^\nu \vH \right )^\dagger \otimes \vD^\mu \vH
      \right )
      - \frac{c_X}{4} g^{\mu \nu}
      \left ( \vD_\rho \vH \right )^\dagger \otimes \vD^\rho \vH \notag \\
      &&&
      +\frac{1}{8} \left (
      \tr{\left ( \vD^\mu \vH \right )^\dagger \vD^\nu \vH}
      -  \frac{c_X}{4} g^{\mu \nu}
      \tr{\left ( \vD_\rho \vH \right )^\dagger \vD^\rho \vH}
      \right )
      \Bigg ] \tau^{aa} \, .
  \end{alignat}
\end{subequations}
Here, $\vH = \frac12 \left(\mathds{1} (v+H) -i w^a \tau^a\right)$, and
$\tau^{aa}$ is the tensor-product representation for the isotensor
case. With those resonances at hand, parameterized simply by their
masses and widths, together with the currents above, one can ingegrate
them out again and derive the corresponding Wilson coefficients of the
dim-8 operators $\mathcal{O}_{S,0}$ and $\mathcal{O}_{S,1}$ in the
section before, for all resonances considered above. The coefficients
\begin{figure}[tb]
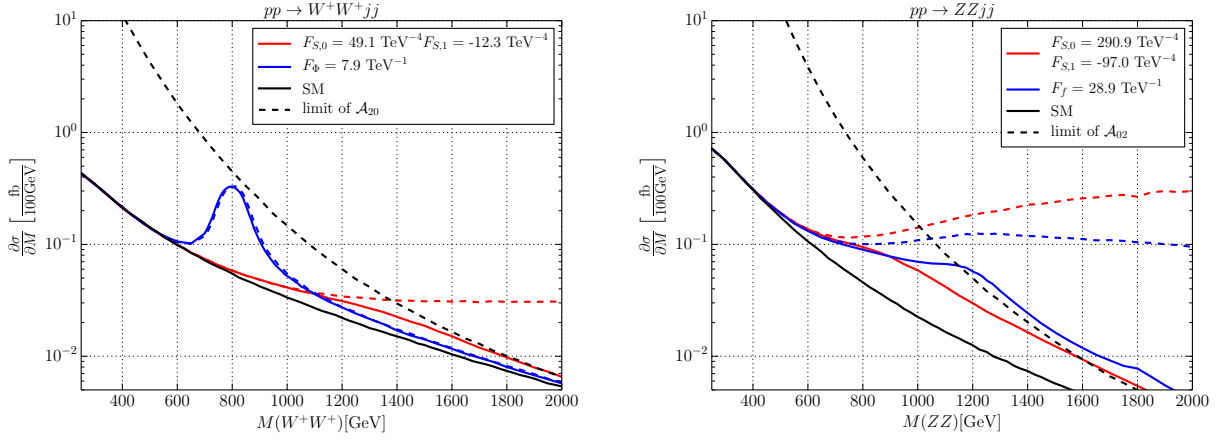

  \centering
  \includegraphics[width=0.48\linewidth]{{{ppWpWp_p_800-0_1dat_pp2}}}
  \includegraphics[width=0.48\linewidth]{{{ppZZ_f_1200-0.4dat_pp2}}}
  \caption{
    \label{fig:isoscalar-tensor}         
    Differential cross sections of a scalar-isotensor resonance (left)
    and an isoscalar-tensor resonance (right). Solid line: unitarized
    results, dashed lines: naive result, 
    black dashed line: limit of saturation of $\mathcal{A}_{22}$ $(W^+W^+)$ /
    $\mathcal{A}_{02}$ $(ZZ)$. Cuts: $M_{jj} > 500$ GeV;
    $\Delta\eta_{jj} > 2.4$; $p^j_T > 20$ GeV; $|\eta_j| > 4.5$. Left:
    $pp\rightarrow W^+W^+jj$, scalar-isotensor
    with $m_\phi= 800 \, \mathrm{GeV}$ and $\Gamma_\phi= 80 \,
    \mathrm{GeV}$, right: $pp\rightarrow ZZjj$, strongly interacting
    isotensor scalar with $m_f= 1200 \, \mathrm{GeV}$ and $\Gamma_f=
    480 \, \mathrm{GeV}$.}
\end{figure}
are listed in the following table, in units of $32 \pi \Gamma/M^5$:
\begin{center}
  \begin{tabular}{c| M{2 cm}  M{2 cm}  M{2 cm} M{2 cm} N}
    & $\sigma$ & $\phi$  & $f$  &  $X$ & \\
    \hline
    $F_{S,0}$ & $\frac{1}{2}$ & $2$ & $15$& 5 &\\[4 ex]
    $F_{S,1}$ & -- & -$\frac{1}{2}$ & -$5$&  -35 &
  \end{tabular}
\end{center}
Tensor resonances as they could arise as Kaluza-Klein recurrences of a
higher-dimensional gravity theory, but also as analogues to tensor
mesons in a composite model, are particularly interesting. They
usually give the largest signal contributions, as here the maximum
number of spin components are involved in the scattering, namely five,
compared to scalar and vector cases. There is a substantial difference
in the theoretical handling of those intrinsic spin degrees of freedom
when dealing with the tensor resonance on-shell and off-shell. In a
full Monte-Carlo simulation (cf. below), one actually simulates the
final state and always has the tensor resonance in off-shell
configurations. Using the analogue of unitarity gauge for tensors, the
propagators lead to a bad high-energy behavior of the amplitudes. Of
course, these could again be treated by a unitarization prescription,
however, it is better to cure most of these issues beforehand. A
symmetric tensor field $f_{\mu\nu}$ has 10 components which are
reduced by the on-shell conditions to five physical components. These
conditions are the tracelessness, $f_\mu^{\phantom{\mu}\mu} = 0$ and
the transversality, $\partial_\mu f^{\mu\nu} = 0$. The original
formulation using the Fierz-Pauli Lagrangian~\cite{Fierz:1939ix} is
not valid off-shell, so we use the St\"uckelberg
mechanism~\cite{Stueckelberg:1900zz,Stueckelberg:1941th} to make the 
off-shell high-energy behavior explicit. Onshell, there is only the
tensor field, $f^{\mu\nu}$, while off-shell there is a vector field,
$A^\mu \sim \partial_\nu f^{\mu\nu}$, which corresponds to the
transversality condition, a scalar implementing the fully contracted
transversality, $\phi \sim \partial_\mu \partial_\nu f^{\mu\nu}$, and
another scalar corresponding to the tracelessness, $\sigma \sim
f_\mu^{\phantom{\mu}\mu}$. By gauge fixing, one of the scalar degrees
of freedom is redundant: $\sigma = - \phi$. The technical details
together with the full Lagrangians and currents for the Fierz-Pauli as
well as the St\"uckelberg picture can be found
in~\cite{Kilian:2014zja}. 

Fig.~\ref{fig:isoscalar-tensor} shows two examples how differential
invariant mass distributions of the diboson system behave at the LHC
in the presence of such resonances. Both plots show different
resonances in different scenarios: the left plot a narrow isotensor
scalar with mass $m_\phi = 800$ GeV and width $\Gamma_f = 80$ GeV, the
right one a strongly-interacting scenario with a broad
isoscalar-tensor resonance of mass $m_f = 1.2$ TeV and width $\Gamma_f
= 480$ GeV. The left plot shows the like-sign $W^+W^+$ channel, the
right one the opposite-sign $W^+W^- \to ZZ$ channel,
respectively. General cuts for selection and signal/background
enhancement are shown in the caption. The full black line is the SM,
the black dashed line shows the corresponding unitarity limit of the
leading partial wave amplitude, the full blue line shows the SM with
the corresponding resonance, while full red line depicts the
approximation with the two Wilson coefficients, $F_{S,0}$ and
$F_{S,1}$. Clearly, if explicit resonances are in the kinematic reach
of the LHC, the EFT is no longer a viable approximation in any
case. Note that even in the simulation with an explicit resonance,
T-matrix unitarization has been applied to unitarize the high-energy
tail of the distribution. As here the amplitudes do have explicitly
complex poles, T-matrix unitarization is actually needed.

\begin{figure}[hp]
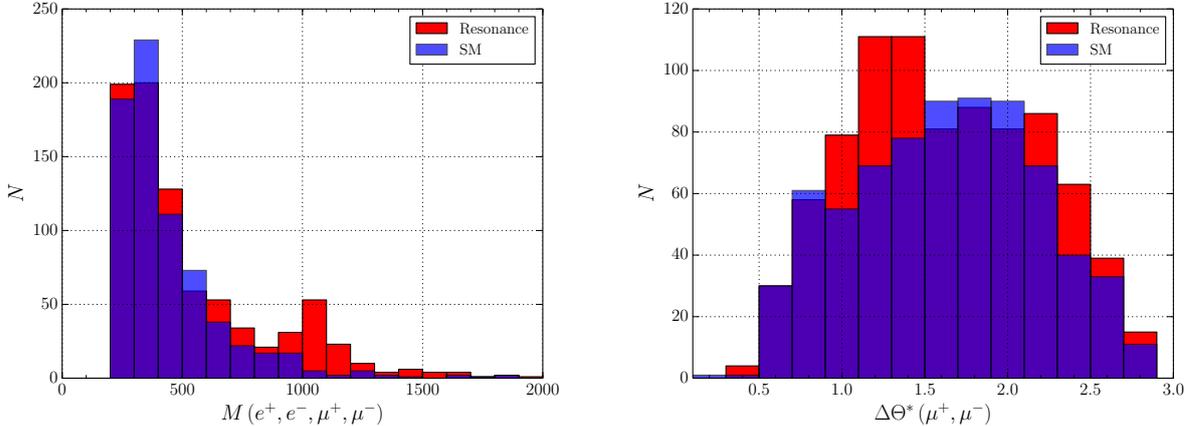

  \centering
  \includegraphics[width=0.48\linewidth]{{{VBS_ppe2mu2_f_r-0.1_1000-mww}}}
  \includegraphics[width=0.48\linewidth]{{{VBS_ppe2mu2_f_r-0.1_1000-theta_star}}}
  \caption{\label{fig:isoscalar_tensor_full}
    Isoscalar-tensor resonance at $m_f=1000$ GeV and
    $\Gamma_f $=100 GeV $pp\rightarrow e^+ e^- \mu^+ \mu^- jj$ at
    $\sqrt{s} = 14 \, \mathrm{TeV}$  with luminosity of $3000 \,
    \mathrm{fb}^{-1}$, with cuts $M_{jj} > 500$ GeV; $\Delta\eta_{jj} > 2.4$;
    $p^j_T > 20$ GeV; $|\eta_j| > 4.5$; $100\; \mathrm{GeV} >
    M_{e^+e^-} > 80\; \mathrm{GeV}$; $100\; \mathrm{GeV} > M_{\mu^+\mu^-} > 80\; \mathrm{GeV}$.
  }
\end{figure}

We have implemented the complete set of dim-6 operators as well as a
complete set of bosonic dim-8 operators together with the prescription
of K-/T-matrix unitarization (for longitudinal VBS) in the Monte Carlo
event generator
\texttt{WHIZARD}~\cite{Kilian:2007gr,Moretti:2001zz}. It contains a
quite elaborate machinery for QCD precision physics, where it uses the
color flow formalism~\cite{Kilian:2012pz}, it has its own parton
shower implementations~\cite{Kilian:2011ka}, and quite recently has
successfully demonstrated its QCD NLO
capabilities~\cite{WHIZARD_NLO}. \texttt{WHIZARD} has been used for a
plethora of BSM studies, and is able to read in external models,
e.g. via~\cite{Christensen:2010wz}. Using this implementation, we
simulated vector boson scattering at the LHC with its design energy of
$\sqrt{s} = 14$ TeV for all kinds of narrower and wider resonances of
different spin and isospin. Fig.~\ref{fig:isoscalar_tensor_full} shows
an example of a isoscalar tensor resonance of mass $m_f = 1$ TeV and a
width of $\Gamma_f = 100$ GeV in the scattering of opposite-sign $W$s
into two $Z$s. A standard set of selection cuts are mentioned in the
caption of the figure. The left plot shows the invariant mass of the
diboson system, which in this case is fully reconstructible, while the
right plot shows the distribution of the opening angle of the two
muons from one of the $Z$s. The latter is one of the angular
observables that could be used to discriminate the spin of such
resonances. More examples can be found in~\cite{Kilian:2014zja}.

\FloatBarrier

\section{Conclusions}

The search for new physics in the electroweak in vector boson
scattering at the LHC can be studied in the context of effective
field theory, however, the introduction of dim-6 and dim-8 operators
leads to a very limited range of applicability of the EFT ansatz. In
many models, dim-8 operators could be the leading contributions
where tree-level effects are forbidden by symmetries, and first
contributions come in at the one-loop level, i.e. at dim-8. LHC as a
hadron collider probes a vast range of energy scales, and
high-energy events tend to (over-)dominate the exclusion limits (or
search potentials) for new models. In most cases this is due to
wrong assumptions on the underlying model, if EFT-based approaches
in regimes are used where perturbative unitarity is lost. We studied
examples of a dim-6 and two dim-8 operators and derived unitarity
limits for the different spin and isospin channels in the scattering
of transverse electroweak vector bosons. Then, a unitarization
method, T-matrix unitarization, that is parameter-free and that is
an extension of the "classic" K-matrix unitarization has been
applied to produce results that are physically meaningful. The
T-matrix unitarization has certain advantages, as it is defined for
amplitudes that are intrinsically complex, and does not rely on the
existence of a perturbative expansion. For weakly coupled amplitudes
without imaginary parts it is identical to K-matrix
unitarization. This procedure is not just an academic exercise, it
allows to produce Monte Carlo events that could actually come from a
quantum field theory realized in nature. Furthermore, it is itself a
possible limit of a limit with a strongly interacting continuum like
in QCD or close to a quasi-conformal fixed point, or it could
correspond to a strongly interacting model right below the onset of
a new resonance that is just a little bit outside the kinematical
reach of LHC. We show examples of cross sections as well as
kinematic and angular distributions to show the effects between
"bare" EFT and unitarized simulations. Beyond this parameter-less
approach to new physics in vector boson scattering, we provided a
set of simplified models taking the SM added by all possible
resonances in the spin-isospin channels to which two EW vector
bosons can couple. We focused on scalar and tensor resonances, while
vector resonances are more complicated due to their potential mixing
with the EW bosons. To account for effects of particularly strongly
interacting models, in addition higher-dimensional operators can be
added. Also, adding just single resonances does not lead to
renormalizable models with sound high-energy behavior, hence, we
also applied T-matrix unitarization to the simplified models. In
order to start with a prescription that already has the best
possible high-energy behavior, we isolated the scalar and vector
degrees of freedom in massive tensor fields via the St\"uckelberg
mechanism to represent explicitly the bad behavior of tensor
propagators in unitarity gauge. We concluded with a fully differential
example for a simplified model with an isoscalar tensor
resonance. Further work will be devoted to the study of transverse $W$
and $Z$ polarizations, the discussion of vector resonances as well as
VBS at future lepton colliders~\cite{Baer:2013cma,Behnke:2013lya}. An
old study of the ILC capabilities~\cite{Beyer:2006hx} will be updated
soon~\cite{CLIC_VV}.

\section*{Acknowledgements}
JRR wants to thank the organizers and particularly Harald Fritzsch
for the invitation, for the local support and the excellent
organization at a great venue. 
  
\bibliographystyle{unsrt}

\end{document}